\documentclass[twocolumn,preprintnumbers,amsmath,amssymb,prb]{revtex4}

\usepackage{graphicx}
\usepackage{dcolumn}
\usepackage{bm}

\begin{document}

\title{Evidence for power-law frequency dependence of intrinsic dielectric response in the CaCu$_{3}$Ti$_{4}$O$_{12}$}

\author{Alexander Tselev}
 \email{tselev@physics.georgetown.edu}
 \altaffiliation[Present address: ]{Department of Physics, Georgetown University, 37th and O St. NW, Washington, DC 20057; also at Institute for Physics of Microstructures, RAS, GSP-105, Nizhny Novgorod, 603950 Russia}

\author{Charles M. Brooks}%
\author{Steven M. Anlage} 
 
\affiliation{%
Center for Superconductivity Research, Department of Physics, University of Maryland, College Park, Maryland 20742-4111
}%
  
\author{Haimei Zheng}%
\author{Lourdes Salamanca-Riba}%

\affiliation{%
Department of Materials and Nuclear Engineering, University of Maryland, College Park, Maryland 20742-2115
}%

\author{R. Ramesh}%
 
\affiliation{%
Department of Materials and Nuclear Engineering and Department of Physics, University of Maryland, College Park, Maryland 20742-2115
}%

\author{M. A. Subramanian}%
 
\affiliation{%
DuPont Central Research and Development, Experimental Station, Wilmington, Delaware 19880-0328
}%


\begin{abstract}
We investigated the dielectric response of CaCu$_3$Ti$_4$O$_{12}$ (CCTO) thin films grown epitaxially on LaAlO$_3$ (001) substrates by Pulsed Laser Deposition (PLD). The dielectric response of the films was found to be strongly dominated by a power-law in frequency, typical of materials with localized hopping charge carriers, in contrast to the Debye-like  response of the bulk material. The film conductivity decreases with annealing in oxygen, and it suggests that oxygen deficit is a cause of the relatively high film conductivity.  With increase of the oxygen content, the room temperature frequency response of the CCTO thin films changes from the response indicating the presence of some relatively low conducting capacitive layers to purely power law, and then towards frequency independent  response  with a relative dielectric constant $\varepsilon'\sim10^2$. The film conductance and dielectric response decrease upon decrease of the temperature with dielectric response being dominated by the power law frequency dependence. Below $\sim$80 K, the dielectric response of the films is frequency independent with $\varepsilon'$ close to $10^2$. The results provide another piece of evidence for an extrinsic, Maxwell-Wagner type, origin of the colossal dielectric response of the bulk CCTO material, connected with electrical inhomogeneity of the bulk material.  
\end{abstract}

\pacs{77.84.-s, 77.22.Ch, 77.55.+f, 81.40.Tv}
\keywords{CaCu$_3$Ti$_4$O$_{12}$, dielectric constant, conductivity, pulsed laser deposition}

\maketitle

\section{Introduction}

There exists considerable interest in the dielectric behavior of the compound CaCu$_3$Ti$_4$O$_{12}$. This material in ceramic and single crystalline forms shows colossal dielectric response with relative permittivity $\varepsilon'$ up to 10$^5$ at room temperature, which is practically frequency independent between DC and 10$^6$ Hz for the temperature range between 100 and 600 K. The dielectric permittivity abruptly drops down to a value $\sim100$ upon lowering the temperature below 100 K. The value $\varepsilon'\sim100$ is also seen at frequencies of a few$\times10^6$~Hz and higher.  \cite{Subramanian:2000,Homes:2001} 
The dielectric response behavior of the material is characteristic of Debye-like relaxation with a single relaxation time. 

Apart from the giant values of the dielectric constant, the independence of dielectric properties with frequency and temperature is a property that is highly desirable for all applications of high-K materials.  The understanding of the mechanism underlying the behavior of CCTO ceramics and single crystals might lead to engineering of new high-K materials with broadly temperature and frequency independent dielectric response, especially those suitable for applications in thin film form. For this reason, CCTO has been a subject of intensive research. However, the origin of the giant dielectric constant of CCTO ceramics and single crystals is so far not understood.  
Attempts were made to explain its behavior as an 'intrinsic' property of the crystal lattice, \cite{Subramanian:2000,Homes:2001} or a property arising due to some 'extrinsic' factors such as lattice defects \cite{RamirezVarmaCondMat:2002} or Schottky barriers at electrode-sample interfaces. \cite{Lunkenheimer:2002} Structural\cite{Subramanian:2000} and spectroscopic \cite{Ramirez:2000,Homes:2001,Kolev:2002,HomesCondMat:2002} investigations of ceramic and single crystal samples along with first principles calculations  \cite{Vanderbilt:2002} suggest an extrinsic, Maxwell-Wagner-type, origin for the dielectric response.
Subramanian \textit{et al.}, \cite{Subramanian:2000} pointed out that non-conducting barrier layers, possibly twin boundaries, separating conducting domains could give rise to the colossal dielectric response of the material through a barrier-layer capacitance mechanism. Similar ideas were expressed also in Ref.~\onlinecite{Vanderbilt:2002}. Based on impedance spectroscopy data, Sinclair \textit{et al.} \cite{Sinclair:2002} later suggested  that CCTO ceramics are a single-step Internal Barrier Layer Capacitor where insulating grain boundaries separate semiconducting grains. 

In this article we report that the dielectric response of the thin epitaxial CCTO films is strongly dominated by a power law, and they behave as semiconductors with hopping conduction. Experiments on oxygen annealing suggest that the films with lower oxygen content are intrinsically more conducting. The most conducting films show the presence of relatively low conducting capacitive layers in their morphology and significantly larger dielectric constants at low frequencies. Finally, CCTO films can be obtained, which show the intrinsic relative permittivity of the compound $\varepsilon'\sim100$ at room temperature.  The observations provide additional evidence and strongly support the arguments for a barrier layer mechanism  of the origin of the colossal dielectric response of CCTO.

\section{Experiment}
\label{sec:Experiment}

CCTO films were grown by PLD using a KrF laser with a wavelength of 248~nm on LaAlO$_3$ (001) (LAO) substrates of $h_s=500$~$\mu$m thickness (relative permittivity $\varepsilon_s=24$) in oxygen atmosphere  from CCTO ceramic targets prepared by DuPont. The average fluence across the laser spot on the target was kept equal to $\sim$2.5~J/cm$^2$. Pulse repetition rates were from 1~Hz to 5~Hz. 
The film composition and structure were characterized by Rutherford Backscattering Spectroscopy (RBS), X-ray diffractometry, Transmission Electron Microscopy (TEM), Atomic Force Microscopy (AFM), and Scanning Near-Field Microwave Microscopy.\cite{Steinhauer:RSI} Electrical properties of the films were investigated using Interdigital Electrode (IDE) capacitors lithographically made on top of the films. We have characterized properties of films of thicknesses $h_f$ ranging from 80 nm to 1500 nm.

\begin{figure}[b]
 \includegraphics[width=8.4cm]{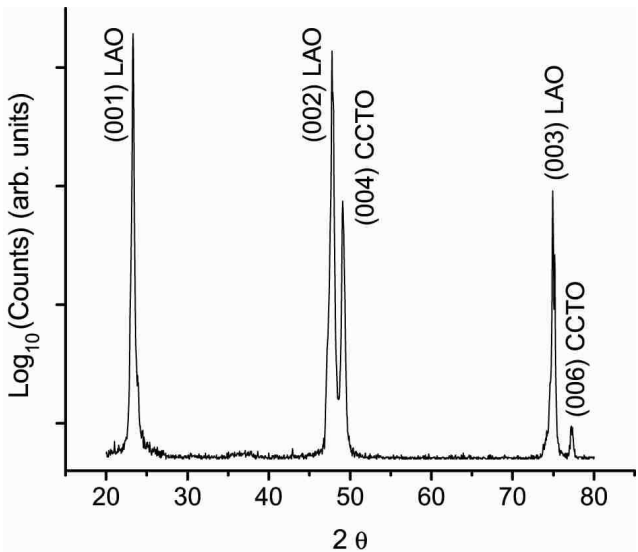}
 \caption{\label{fig:Xray}
  Typical X-ray diffraction $\theta$-2$\theta$ pattern of an epitaxial CCTO film on (001) LaAlO$_3$ substrate studied in the present work. 
}
\end{figure}

The following growth conditions were found optimal for both crystalline quality of the films and roughness of the film surface: Substrate temperature is 695~$^\circ$C, oxygen pressure is 60--200~mTorr, substrate is slightly immersed in the visible plasma plume. The deposition rate for these conditions is 0.6--1~\AA/pulse. After deposition, samples were cooled down at the rate of 5~$^{\circ}$C/min in oxygen at pressure $P_{\text{O}_2}\approx750$~Torr.
Films produced under these conditions were found to be rather conducting. Their room temperature conductivity varied in the range 10$^{-7}$--10$^{-4}$~$\Omega^{-1}$cm$^{-1}$. To make more insulating films, we decreased the deposition rate of the films by decreasing the laser repetition rate and/or annealed films in oxygen after deposition. Post-deposition anneals were carried out in flowing oxygen at atmospheric pressure and at a temperature of 695~$^{\circ}$C. 

The conductance of as-deposited films was found to be strongly dependent on deposition conditions and poorly reproducible. Increase of the background oxygen pressure during deposition resulted in less conducting films. Post-deposition annealing in oxygen always decreased the film conductance and created films with similar properties.  High-Resolution TEM investigations of the CCTO/LAO interface show a very sharp and clean interface indicating that there is no chemical reaction between the film and substrate during annealing. Table~\ref{tab:parameters} lists parameters of five samples chosen based on their room temperature conductivity and discussed in the paper.  

\begin{table*}
\caption{\label{tab:parameters}Parameters of CCTO thin film samples discussed in the paper: $P_{\text {O}_2}$ is the oxygen pressure in the chamber during deposition, $t_{\text {ann}}$ is the duration of annealing in oxygen, $h_f$ is the nominal film thickness, $\sigma_{\text {DC}}$ is the film DC conductivity at room temperature, $n$ is the number of fingers of the interdigital capacitor, 'electrodes' is the electrode material.}
\begin{ruledtabular}
\begin{tabular}{ccccccc}
sample &$P_{\text {O}_2}$ (mTorr)&$t_{\text {ann}}$ (hrs)& $h_f$ (nm)& $\sigma_{\text {DC}}$ ($\Omega^{-1}{\text {cm}}^{-1}$) &
 $n$ & electrodes\\
\hline
1& 80 & -- & 230 & $3\times10^{-4}$ & 100 & Pt \\
2& 200 & -- & 1500 & $1\times10^{-6}$ & 50 & Pt\\
3& 100 & 4 & 230 & $2\times10^{-7}$ & 50 & Pt/Au\\
4& 60 & 15 & 430 & $2\times10^{-8}$ & 50 & Pt\\
5& 80 & 32 & 230 & $2\times10^{-10}$ & 100 & Pt\\
\end{tabular}
\end{ruledtabular}
\end{table*}

CCTO grows epitaxially with the (00\textit{k}) direction perpendicular to the substrate surface with a high quality of epitaxy, in agreement with the behavior reported in the literature.\cite{Si:2002,LinChen:2002}   Figure~\ref{fig:Xray} shows a typical result of X-ray diffraction $\theta$--$2\theta$ scan of a CCTO film. 
The full width at half-maximum of the X-ray diffraction rocking curves around the (004) CCTO peak of as-deposited films was 0.57--0.60$^\circ$. RBS studies of the films showed cation ratios corresponding to that of CCTO. TEM studies showed that the film-substrate lattice mismatch was relaxed by misfit dislocations at the interface.

\begin{figure}[h]
 \includegraphics[width=7.5cm]{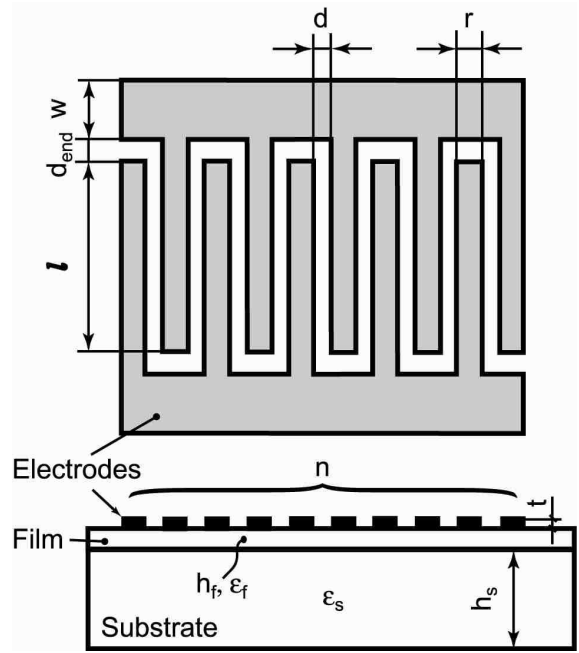}
 \caption{\label{fig:capacitor}
Layout of an interdigital capacitor used to measure the dielectric and conducting properties of CCTO films. Parameters are defined in the text.
}
\end{figure}

Interdigital capacitors were made on top of the films after all film fabrication steps. The layout of an interdigital capacitor and notation of its dimensions are shown in Fig.~\ref{fig:capacitor}. The capacitor fingers were of $r=25$~$\mu$m width and $l=700$~$\mu$m length and were separated by gaps of width $d=15$~$\mu$m. The electrodes consisted either of $\sim$20~nm thick Pt layer or a $\sim$10~nm thick Pt adhesion layer followed by a $\sim$20~nm thick Au layer and were deposited through a photoresist mask by PLD at room temperature with following lift-off. The parasitic resistance of  the pure Pt electrodes is much larger than that of Pt/Au electrodes and  reaches 180 $\Omega$ for some samples. The parasitic resistance of Pt/Au electrodes does not exceed 15~$\Omega$. The impedance of the IDE capacitors was measured by means of LCR meters HP~4284A (20 Hz -- 1 MHz), Agilent 4285A (75 kHz -- 30 MHz), and SRS 510 (100 Hz -- 100 kHz). 
 
	If the film thickness $h_f$ is much smaller than the interdigital capacitor dimensions (see Fig.~\ref{fig:capacitor}): $h_f \ll r$, $d$, $d_{\text{end}}$, $h_s$, $w$, $l$,  the number of fingers $n \gg 3$, and thickness of capacitor electrodes $t \ll r$, which was true in our experiments, the following approximate formula can be used to extract the film relative permittivity $\varepsilon'$ from the measured capacitance $C_{\text {meas}}$ and the capacitance $C_s$ of the same IDE capacitor made on a bare substrate with a relative permittivity $\varepsilon_s$:
	\begin{equation}
	\varepsilon'\approx\varepsilon_s+\dfrac{C_{\text {meas}}-C_s}{C'}~, \label{eq:epsilon}
\end{equation}
where 
	\begin{equation}
	C'=\dfrac{1}{2}(n-3)\varepsilon_0 l q_n + 2\varepsilon_0 l q_3 + n(2+\pi)\varepsilon_0 r q_{\text {end}}\,.
	\end{equation}
In the last expression, $\varepsilon_0$ is the permittivity of vacuum, and the coefficients are: 
\begin{eqnarray}
	q_n& = &\dfrac{\pi}{3 \ln 2 + \pi d/(2h_f)}\\
	q_3& = &\dfrac{\pi}{4 \ln 2 + \pi d/h_f}\\
	q_{\text{end}}& = &\dfrac{\pi}{4 \ln 2 + \pi d_{\text{end}}/h_f}\,. \label{eq:qs}
\end{eqnarray}
Eqs.~(\ref{eq:epsilon})-(\ref{eq:qs}) follow from expressions for capacitance of IDE capacitors on multilayered substrates derived in Ref.~\onlinecite{Gevorgian:1996} and approximate the exact formulas with an accuracy of  $\sim$1\% under our conditions. 

The conductance of the LAO substrate is much smaller than that of the CCTO films as was determined from the loss tangent measurements of the substrate material. The frequency dispersion of the substrate permittivity is negligibly small in the whole frequency range, and the substrate contribution to the frequency dependent part of the capacitance can be neglected.

\section{Results and Analysis}
\label{sec:ResultsAndDiscussion}

\subsection{Room temperature frequency dependence of dielectric response}

\begin{figure*}[t]
\includegraphics[width=17.4cm]{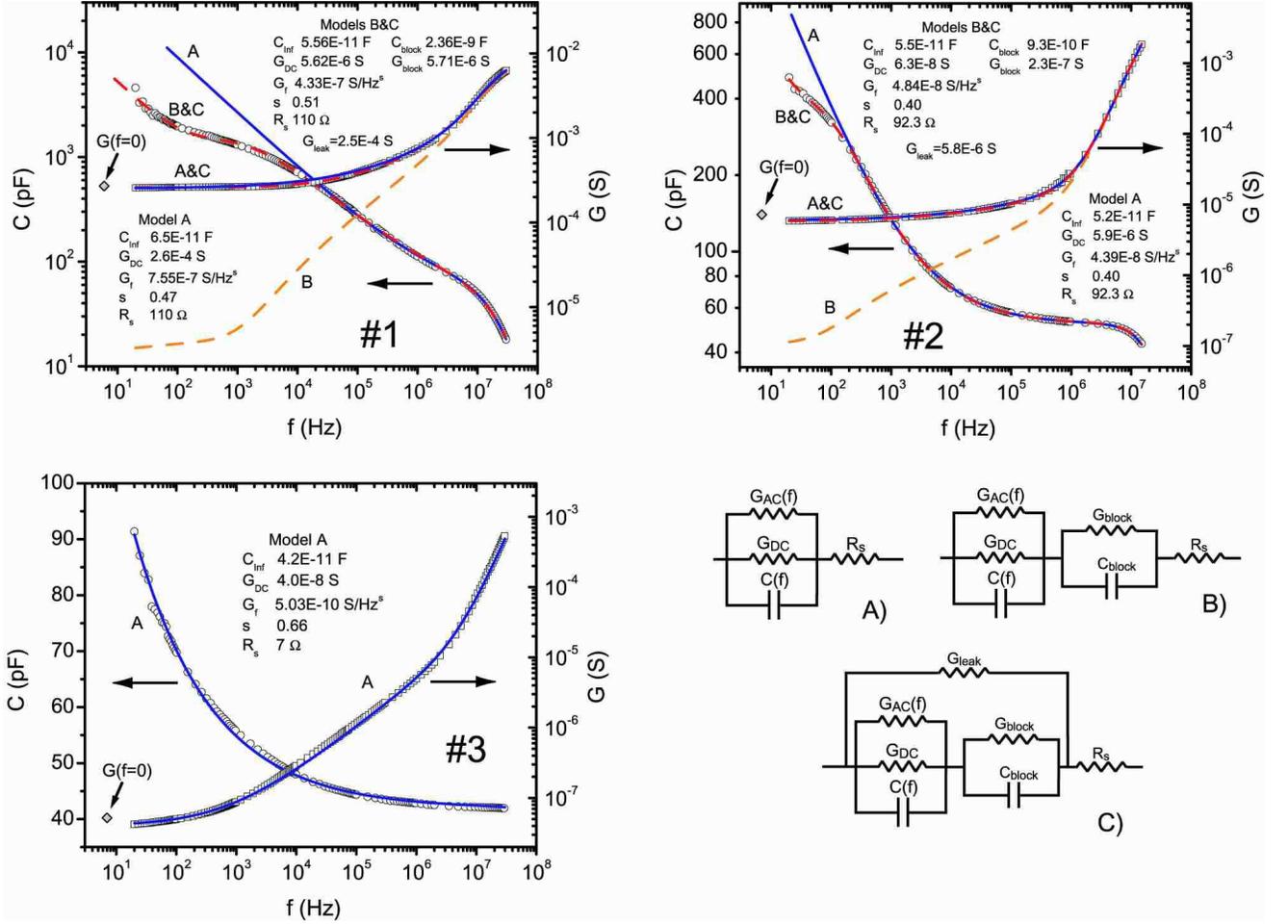}
\caption{\label{fig:FrequencyResponse}
  Room temperature frequency response of three IDE capacitors on the CCTO films listed in Table~\ref{tab:parameters}. Symbols are the experimental data, solid and dashed curves are fittings to the models shown in the lower right panel with AC conductance $G_{\text{AC}}(f)=G_f f^s$ and capacitance $C(f)=C_{\infty}+(2\pi)^{-1} G_f\,{\text {tan}}(\pi s/2)\, f^{(s-1)}$ as follows from the universal dielectric response law.\cite{Jonscher:Nature:1977} $R_s$ is the parasitic resistance of the capacitor electrodes. Parameters of the fits are shown in corresponding panels. The diamonds show the DC conductance of the capacitors $G(f=0)$.}
\end{figure*}

Figure~\ref{fig:FrequencyResponse} shows the room temperature frequency dependence of the capacitance $C$ and the conductance $G$ measured in the CGp representation (parallel C and G) from 20 Hz to 30 MHz for the three most conducting samples listed in Table~\ref{tab:parameters}. The open symbols are experimental data and the solid and dashed lines are fits for the measurement data using the capacitor models shown in the lower right panel  of Fig.~\ref{fig:FrequencyResponse}. As is seen, the experimental data for the CCTO thin films do not feature the clear Debye-like behavior demonstrated by CCTO ceramics and single crystals. Instead, their dielectric behavior has a strong contribution of a power law. In a log-log representation, the power law is linearized and it can be recognized in the capacitance data for samples 1 and 2, as well as in the conductance data for  sample~3. 

To analyze details of the dielectric response of our CCTO films, we start with the simplest equivalent circuit model taking into account the power law frequency behavior of the capacitors, model A in Fig.~\ref{fig:FrequencyResponse}. This model includes a capacitance $C(f)$ describing the real part of the relative permittivity of the material, conductances $G_{\text{DC}}$  and $G_{\text{AC}}(f)$ describing the intrinsic DC conductance and the AC conductance of the material,  and the parasitic resistance of the capacitor electrodes $R_s$. 
The AC conductance $G_{\text{AC}}(f)$ and the capacitance $C(f)$ of the model follow the power law (Curie-von Schweidler law, or the universal dielectric response (UDR) law \cite{Jonscher:Nature:1977,Jonscher:Book,Lunkenheimer:2002}): 
\begin{equation}
  G_{\text{AC}}(f)=G_f f^s,
  \label{eq:G}
\end{equation}  
\begin{equation}
  C(f)=C_{\infty}+(2\pi)^{-1} G_f\,{\text {tan}}(\pi s/2)\, f^{(s-1)}.
  \label{eq:C}
\end{equation}  

The conductance data of all the three samples can be well fit with this model in the whole frequency range, whereas the capacitance data for samples 1 and 2 can be fit only partially, as shown by solid lines in Fig.~\ref{fig:FrequencyResponse}.  The parameters of the fittings are displayed in the corresponding panels. The capacitance data for samples 1 and 2 deviate from the UDR fits at low frequencies, which might indicate a contribution from possible capacitive layers with lower conductivity (blocking layers).

Therefore, we further continue with the analysis of the responses of samples 1 and 2 only.
To take the low frequency deviation into account, we add a parallel RC-circuit  with frequency independent parameters in series with elements of model A as shown by model B in Fig.~\ref{fig:FrequencyResponse}. Now, very good fits can be obtained for the capacitance response of both samples 1 and 2 in the whole frequency range, however the corresponding conductivity curves significantly deviate from the experimental points (see dashed lines denoted with the letter B in the upper panels of Fig.~\ref{fig:FrequencyResponse}). Addition of one more element, a frequency independent leakage conductance ${\text G}_{\text {leak}}$ around the UDR and the blocking components, as shown in model C Fig.~\ref{fig:FrequencyResponse}, makes it possible to fit simultaneously both capacitance and conductance data with one set of parameters (see dashed lines denoted with the letter C in the upper panels of Fig.~\ref{fig:FrequencyResponse}; the conductivity curves for models A and C almost overlap).

We would like to note here that model B both with UDR-dependent and with frequency independent parameters was invoked in the literature to explain the colossal dielectric response of the bulk CCTO. \cite{Lunkenheimer:2002,Sinclair:2002,SinclairAdvMater:2002} For a pronounced Debye-like behavior, as demonstrated by bulk CCTO,  it is necessary that $G_{\text {block}}\ll G_{\text {DC}}$. Note that a capacitor in series with an RC circuit represents systems with Debye dipolar relaxation.\cite{Jonscher:Book} 

The frequency response of sample 3 is different from that of samples 1 and 2. Model A is sufficient to fit the response of sample 3 (see solid lines in the lower left part of Fig.~\ref{fig:FrequencyResponse}). Moreover, modes B and C cannot satisfactory describe the response of  sample 3. An attempt to fit the capacitance data with the use of model C resulted in a very large $C_{\infty}\approx59$~pF, which gives $\varepsilon'_{\infty}\approx460$, an unrealistically high value. The fit with the use of model A results in  $C_{\infty}\approx42$ pF and $\varepsilon'_{\infty}=90$. 

As can be concluded, sample 3 is either missing any insulating blocking layer components that are present in samples 1 and 2 and in bulk materials, or their effect is removed because the intrinsic conductivity of the material dropped below some critical value, and we observe the intrinsic frequency response of the material in the whole frequency range.

The values of $C_{\infty}$ found from the fit for sample 1 using model C  is lower than the capacitance of the same IDE capacitor on a bare substrate, which was attributed to a high leakage in the film.  The fit for sample 2 resulted in  $\varepsilon'_{\infty}=64\pm20$. The values of $\varepsilon'_{\infty}$ for samples 2 and 3 are close to $\sim$80 found at terahertz frequencies in Ref.~\onlinecite{Homes:2001} for the dielectric constant of CCTO single crystals. Additionally, measurements at 1.4~GHz made with a microwave microscope\cite{Steinhauer:RSI} on sample 3 gave $\varepsilon'=40\pm20$. Application of up to $\pm$40~V bias voltage to the capacitors did not change the dielectric response of the films, demonstrating the linearity of the dielectric properties of the CCTO films. The parasitic resistance of the capacitor electrodes $R_s$ causes a downturn in the capacitance as well as a strong upturn in the conductance data upon increase of frequency at the highest frequencies of the range. These features are more pronounced for samples with larger $R_s$ as seen in Fig.~\ref{fig:FrequencyResponse}. The difference in $R_s$ of the three samples arises from different electrode materials and different thicknesses of the Pt film. The values of the parameter $R_s$ obtained from the fittings of Fig.~\ref{fig:FrequencyResponse} are consistent with the resistivity of the Au and Pt thin films and with the geometry of the capacitor electrodes.

The capacitance of sample 5 (Table~\ref{tab:parameters}), obtained with a long anneal in oxygen, is weakly frequency dependent at room temperature and  remains close to a value corresponding to a CCTO film permittivity of $\sim100$ in the whole frequency range.

In the present state of our knowledge of CCTO and within the scope of the present study, we cannot identify the nature of conduction in general and the leakage in particular in our CCTO films. CCTO is a Mott-Hubbard insulator\cite{Vanderbilt:2002} and might become conducting by relatively small deviations from stoichiometry, by doping with oxygen vacancies, or due to defects.  A possible scenario consistent with model C can include leakage along grain boundaries due  to deviation of stoichiometry and accumulation of defects along them. Then, as follows from model C,  the electrodes do not create barriers at interfaces with the leaking component, and the blocking layers should be internal to the less conducting grains. 
Annealing in oxygen leads to restoration of defects and to oxidation of the material inside the grains and in the grain boundaries. Both these processes result in drop of intrinsic and leakage conductance of the films.

\subsection{Temperature dependence of permittivity and conductivity}

\begin{figure}[h]
 \includegraphics[width=8.2cm]{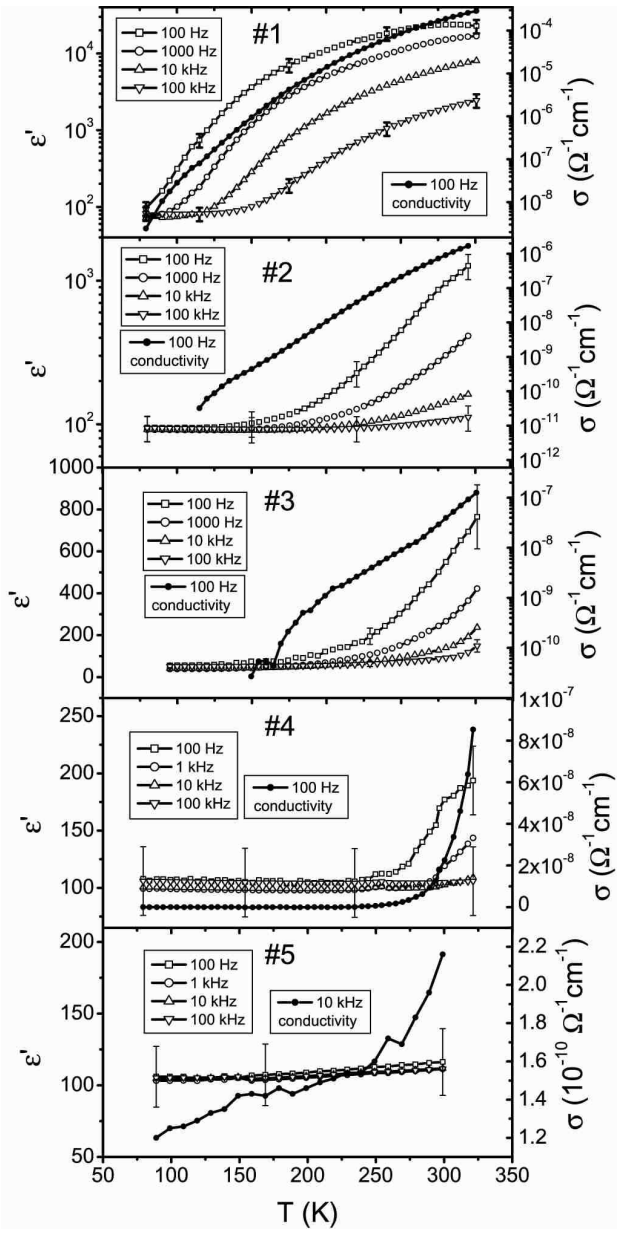}
 \caption{\label{fig:TemperatureDependence}
  Temperature dependence of dielectric response for CCTO films listed in Table~\ref{tab:parameters} between 80 K and 300 K. $\varepsilon'$ is the real part of the relative permittivity shown for four frequencies (open symbols): 100 Hz, 1000 Hz, 10 kHz, and 100 kHz. $\sigma$ is the film conductivity at a frequency of 100~Hz.}
\end{figure}

Figure~\ref{fig:TemperatureDependence} shows the temperature dependence of the film relative permittivity at four frequencies logarithmically separated by one decade, and the 100 Hz conductivity for all the five samples listed in Table~\ref{tab:parameters} in the temperature range from 80 to 300 K. The 100 Hz conductivity was obtained from the capacitor loss tangent data. Similar to the bulk materials \cite{Subramanian:2000,Homes:2001} and to the thin films of Ref.~\onlinecite{Si:2002}, the permittivity of the CCTO films starts to rise from a value of about 100 with increasing temperature. At the lowest temperatures, the permittivity is independent of frequency within the measurement error. For the sample with the lowest conductivity, the relative permittivity is close to 100 almost independent of temperature and frequency. 

\begin{figure}[t]
 \includegraphics[width=6.8cm]{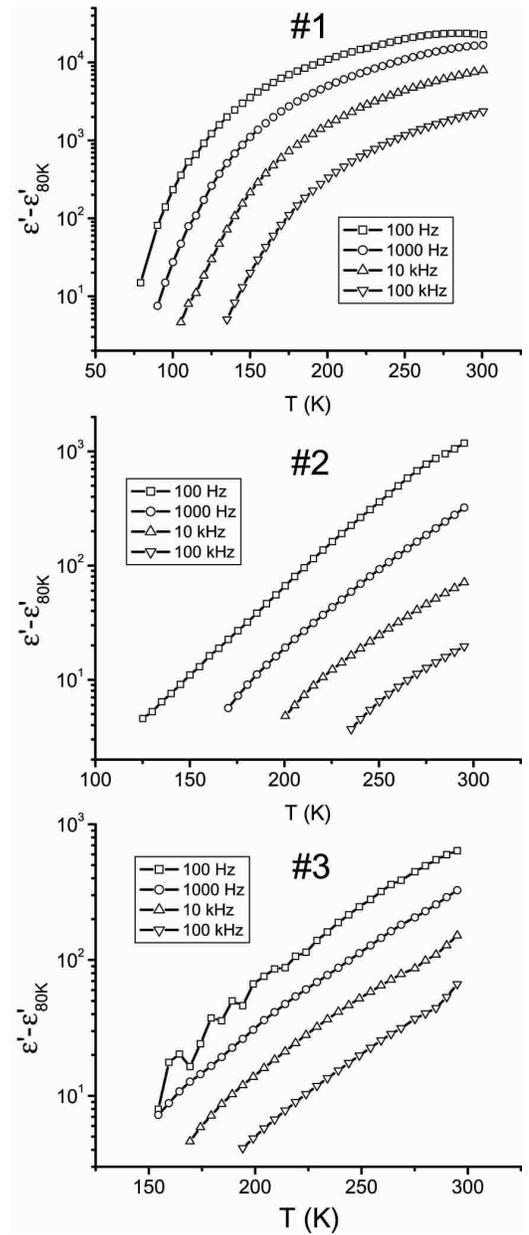}
 \caption{\label{fig:ParalLines}
 The permittivity temperature dependence data of Fig.~\ref{fig:TemperatureDependence} are re-plotted here for the three most conducting samples after subtraction of  $\varepsilon'_{\text {80K}}$. Only data points with significant signal-to-noise ratio are shown. In the case of sample 1, the value of $\varepsilon'_{\text {80K}}$ for 1000 Hz was used to plot the curve for 100 Hz.
}
\end{figure}

To find out whether the power-law behavior found at room temperature is preserved at lower temperatures, we re-plotted the permittivity data of Fig.~\ref{fig:TemperatureDependence} in the following way. We subtracted  values of the relative dielectric constant obtained at a temperature of 80 K $\varepsilon'_{\text {80K}}$ from the data points for each frequency and plotted the difference ($\varepsilon'-\varepsilon'_{\text {80K}}$)  on a log-scale versus temperature. We take into account here that $\varepsilon'_{\text {80K}}\approx \varepsilon'_{\infty}$ obtained at room temperature. The result is displayed in Fig.~\ref{fig:ParalLines}. Only data points with a significant signal-to-noise ratio are shown. In the case of sample 1, the value of $\varepsilon'_{\text {80K}}$ for 1000~Hz was used to plot the 100 Hz curve. 

We note that the power law is linearized in the log-log representation. Since the four frequencies used for the plots are logarithmically spaced by one decade, this representation immediately shows how close to the UDR the frequency dependence of the permittivity remains upon lowering the temperature.  For instance, the curves for sample 3 look like four equally separated parallel lines, which indicates preservation of the UDR frequency response up to at least 170 K with a small temperature dependence of the $s$-parameter. 

\begin{figure}[t]
 \includegraphics[width=7.8cm]{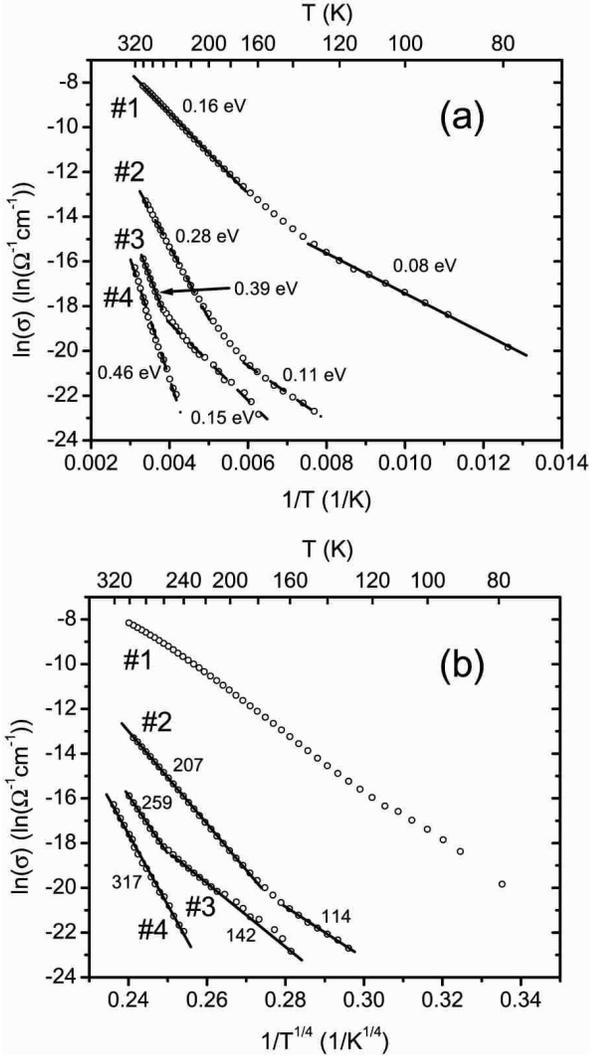}
 \caption{\label{fig:Arrhenius}
(a)~Arrhenius plots for 100 Hz conductivity of the four most conducting samples listed in Table~\ref{tab:parameters}. Dashed lines show the parts of the curves that can be approximately linearized by the Arrhenus representation. The numbers are activation energies. 
(b)~Conductivity at 100 Hz of the four most conducting samples of Table~\ref{tab:parameters} shown in the representation linearizing the law  $\sigma\sim{\text {exp}}(-B/T^{1/4})$. Piecewise, the conductivity data for samples 2--4 can be better linearized in this representation than in the Arrhenius representation. Numbers are the values of the parameter $B$ in units of ${\text {K}^{1/4}}$.
}
\end{figure}

Figure \ref{fig:Arrhenius}(a) shows Arrhenius plots for the 100 Hz conductivity of samples 1--4. According to the fits and model C of Fig.~\ref{fig:FrequencyResponse}, the conductance  of the films is strongly dominated by the highly conducting leakage component in the films. Nevertheless, the analysis of temperature dependence of the film conductivity can provide some insight into the conduction mechanisms of CCTO. We assume here that the contribution of the AC part of the UDR conductivity is negligibly small at this frequency, as can be deduced from the room temperature data of Fig.~\ref{fig:FrequencyResponse}. The Arrhenius representation linearizes the thermally activated conduction $\sigma\sim {\text {exp}}(-E_a/k_{\text B}T)$. As can be seen after careful consideration, only the data for sample 1 from  80 K to about 130 K and from 170 K to 300 K can be well linearized in this representation, with activation energies of 0.08 eV and 0.16 eV, correspondingly. The behavior of the three other curves deviates from the thermally activated behavior. However, the three curves can be piecewise linearized in the representation ${\text {ln}}\sigma$ versus $1/T^{1/4}$, as shown in Fig. \ref{fig:Arrhenius}(b). The latter representation corresponds to the 3D variable-range hopping conduction model\cite{book:MottDavis} with $\sigma\sim{\text {exp}}(-B/T^{1/4})$. The data for sample 2 can be linearized in this representation in a large temperature range from 180 K to room temperature. The data for sample 3 can be better linearized in the representation ${\text {ln}}\sigma$ versus $1/T^{1/4}$ only for lower temperatures. At higher temperatures, both the representation are indistinguishable. And the data for sample 4 can be linearized in the latter representation in the whole temperature range, where the conductivity is measurable, i. e., between about 240 K and room temperature. This analysis suggests that the hoping of localized charge carriers is the dominant mechanism of charge transport in the samples with lower conductivity, whereas thermally activated transport prevails in the conduction of sample~1. 

\subsection{Analysis and Discussion}

At this stage we can make a preliminary assessment of our results. The most striking feature of the epitaxial CCTO thin films is that their dielectric response is strongly dominated by a power-law in frequency, in contrast to the Debye-like  response of the bulk material. Experiments on annealing of the films in oxygen suggest that films with lower oxygen content are intrinsically more conducting. With increase of the oxygen content the room temperature frequency response of the CCTO thin films evolves from the response indicating presence of some relatively low conducting capacitive layers to purely power law, and then towards frequency independent  response  with $\varepsilon'$ close to 100. The most conducting films show significantly larger room temperature dielectric constants at low frequencies ($\varepsilon'\sim3\cdot10^4$ at $f=20$~Hz).

The power-law frequency dependence of dielectric properties is a behavior typical for semiconducting materials with hopping localized charge carriers.\cite{book:MottDavis,Jonscher:Nature:1977,Jonscher:Book,Lunkenheimer:2002,Bobnar:2002}
This result, along with the observed strong correlation between the temperature behavior of conductivity and permittivity, allow us to conclude that we are dealing with carrier polarization in the CCTO films. The effect of annealing in oxygen on conductivity suggests that oxygen deficit is a cause of the relatively high film conductivity.  
  
The observations described above strongly support arguments in the literature  \cite{Subramanian:2000,Vanderbilt:2002,Lunkenheimer:2002,Sinclair:2002,SinclairAdvMater:2002} on the role of blocking barrier layers in CCTO and evidence for an extrinsic, Maxwell-Wagner type, origin of the colossal dielectric response of CCTO ceramics and single crystals. They also imply that the blocking layers must be insulating, thus cutting down on the number of possible microscopic morphologies suggested for bulk CCTO by Cohen {\it et al.}\cite{Cohen:2003}  We did not find any significant dependence of film properties on thickness, in contrast to Ref.~\onlinecite{Si:2002}.

It is of interest in this context to directly compare the characteristics of the dielectric responses of thin films and some of the CCTO bulk samples. The room temperature measurement of the response of the ceramic target we used for film deposition (not shown) yielded  $\varepsilon'_{\text {100Hz}}\approx4000$ and a pronounced Debye-like response. The fit of the response with model B, Fig.~\ref{fig:FrequencyResponse}, resulted  in 
${\text G}_{\text {DC}}/{\text G}_{\text {block}}=5\cdot10^4$ with an intrinsic DC conductivity of the ceramic $\sigma_{\text {DC}}^{\text {int}}\approx3\cdot10^{-4}$ $\Omega^{-1}{\text {cm}}^{-1}$, assuming that the blocking layers are thin and constitute a small fraction of the sample volume. For the thin film sample 1,  ${\text G}_{\text {DC}}/{\text G}_{\text {block}}\approx1$ and the internal conductivity $\sigma_{\text {DC}}^{\text {int}}\approx7\cdot10^{-6}$ $\Omega^{-1}{\text {cm}}^{-1}$, again assuming that blocking layers along with leakage paths make up only a small fraction of the sample volume. A comparison of these values suggests that the ceramic material is intrinsically more conducting and its barrier layers block conductivity more efficiently.

It is still an open question as to what microscopic feature(s) produce the blocking layers in CCTO samples. Lunkenheimer {\it et al.}\cite{Lunkenheimer:CondMat:2004}  recently showed that contacts  contribute to the dielectric response of the CCTO bulk ceramic samples, and this contribution can constitute a significant fraction of their response at low frequencies. However, to date there is no unambiguous experiment that demonstrates that the influence of electrodes alone can explain all the behavior of CCTO dielectric response especially in single crystals. Apart from the contribution of electrode-sample interfaces, a microscopic picture of hopping localized charge carriers confined within small domains, whose boundaries make up large energy barriers for hopping, remains plausible. Within this picture, the boundaries of the small domains constitute a dense network without ``holes'' for conductivity, and the exact morphology and characteristic scales of the network such as domain size are dependent on sample processing conditions. Suggestions for possible domain boundaries include grain boundaries,\cite{Sinclair:2002,SinclairAdvMater:2002} twin boundaries,\cite{Subramanian:2000} compositional  and anti-phase domain boundaries.\cite{Cohen:2003}   We take this opportunity to examine the microstructure of our films for evidence of (or absence of) any of the proposed blocking layers with focus on possible twinning of the films. The films showing a dielectric behavior suggestive of the presence of some barrier layers, similar to sample~1 discussed above, are of particular interest.  

The films are epitaxial, but morphologically they consist of grains separated by grain boundaries. Taking into account the results of fits of the room temperature dielectric response (Fig.~\ref{fig:FrequencyResponse}), most probably the grain boundaries in the films are more conducting than the grains and cannot act as barrier layers. However, one cannot exclude grain boundaries as possible blocking layers in CCTO bulk ceramics based on this fact, because the microstructure of grain boundaries can be different in thin films and ceramics due to significantly different fabrication processes. 

\begin{figure}[b]
 \includegraphics[width=8.2cm]{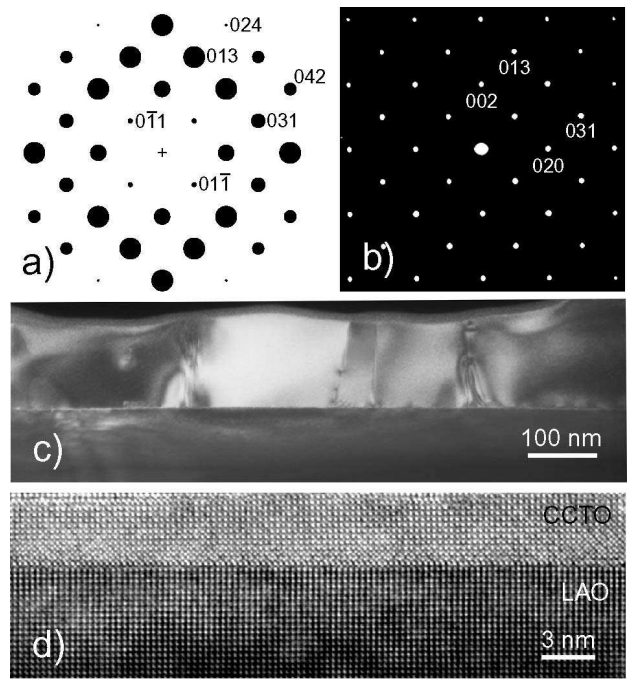}
 \caption{\label{fig:TEM}
  (a) Simulated electron diffraction pattern corresponding to ideal, twin-free CCTO lattice with symmetry group Im\={3}; the spot diameters are proportional to their intensity. 
  (b)~Small-area diffraction pattern taken from the CCTO film. 
  (c) TEM Dark Field image of a cross-section of a CCTO film on LAO; the image is taken using the (013) spot of the film pattern. 
  (d)~HRTEM image of the film-substrate interface of a sample annealed 15~h at 695~$^{\circ}$C. 
}
\end{figure}

	To investigate the microstructure of our films, we have performed detailed plan-view and cross-sectional TEM study of the films. Figure~\ref{fig:TEM}(a) shows a simulated electron diffraction pattern from the ideal, twin-free, CCTO lattice with the Im\={3} space symmetry group. Note that due to the Im\={3} symmetry of the lattice the diffraction pattern is not symmetric relative to the $<$0{\it ll}$>$ directions of the pattern. Namely, pairs of spots within a set, e.g. sets \{013\} or \{024\}, located symmetrically relative to the $<$0{\it ll}$>$ directions have significantly different intensities. As reported in Ref.~\onlinecite{Subramanian:2000}, the Laue group of  twinned CCTO single crystals appeared to be  m\={3}m instead of m\={3}. This means that a CCTO lattice containing twins would produce a diffraction pattern where all spots within each set \{013\}, \{024\}, etc., are of equal intensity. The spot positions would be the same for both m\={3} and m\={3}m Laue groups.  The experimental diffraction pattern in Fig.~\ref{fig:TEM}(b) does not show the asymmetry in the intensities of the spots seen in the simulated diffraction pattern in Fig.~\ref{fig:TEM}(a).  At the same time,  the relative intensities of the spots in the experimental diffraction pattern are different from those expected for the twinned lattice. A possible explanation for the lack of asymmetry in the intensity of the spots is that the CCTO film is twinned at a scale smaller than the area that the diffraction pattern came from and, since the film consists of slightly misoriented grains, the experimental diffraction pattern of Fig.~\ref{fig:TEM}(b) does not posses either a clear m\={3} or m\={3}m symmetry. It is possible also that the films are not twinned but the spots in the diffraction pattern have approximately the same intensity due to multiple diffraction. Microdiffraction patterns obtained with a beam of 40 nm in diameter did not show the asymmetric patterns either, indicating that if there are twins in the crystal they must be smaller than 40 nm in size. 
	
To further identify twins in the CCTO films, we implemented dark field imaging  with use of diffracted beams corresponding to spots of sets \{013\} and \{024\}. Twins contributing differently to spots of these sets should give rise to twining contrast in dark field images. Figure~\ref{fig:TEM}(c) shows a dark field image of the CCTO film cross-section taken with the (013) spot. We do observe contrast between grains here. This contrast can be due to slight misorientation between grains and because the grains can be ``twins'' to each other. Some contrast observed within grains is most probably caused by strain. There is no contrast seen within grains that is consistent with the presence of twinning. Using the highest possible magnification accessible with this technique, we are able to state that the films do not contain twin domains larger than 10 nm. The presence of smaller twins cannot be excluded, however. It is believed that the twins are in the Ti-O sublattice, making their observation by high-resolution TEM difficult. \footnote{Straightforward HRTEM does not reveal any twinning.  Modeling of HRTEM images with varying defocus and sample thickness for the two CCTO structures show that there exist certain features which could in principle allow one to distinguish  between twinned and untwinned states in the case of nano-twins.}

\section{Conclusions}
\label{sec:Conclusions}

In conclusion, CaCu$_3$Ti$_4$O$_{12}$ thin films grown epitaxially on LaAlO$_3$ (001) substrates by PLD were found to be semiconducting with frequency response strongly dominated by a power-law, typical of materials with localized hopping charge carriers. 
The room temperature frequency response of the CCTO thin films converges with increase of the oxygen content from a response indicating the presence of the relatively low conducting capacitive layers, through purely power law, towards frequency independent  response  with $\varepsilon'$ close to 100. Films with lower oxygen content are intrinsically more conducting. Lowering the temperature leads to decrease of the film conductance and to decrease of the dielectric response, which remains dominated by the power law. Below $T\sim80$ K, the dielectric response of the films was found to be frequency independent with $\varepsilon'$ close to 100. The observations suggest that we are dealing with carrier polarization in the CCTO films. The results can be considered as additional evidence that the giant dielectric constants of bulk and thin films of CCTO reported in the literature are of extrinsic origin of Maxwell-Wagner type. Detailed plan-view and cross-sectional TEM studies did not reveal twin domains of size larger than 10 nm in our CCTO films.

\section{Acknowledgements}
\label{sec:Acknowledgements}

We would like to thank J. B. Neaton for valuable discussions, R. P. Sharma and V. Kulkarni for performing RBS measurements. The work was supported by the University of Maryland/Rutgers NSF MRSEC under grant DMR-00-80008.


\end{document}